\documentclass{article}
\usepackage{ltwol2e}

\def\Journal#1#2#3#4{{#1} {\bf #2}, #3 (#4)}

\def\NPB{{\em Nucl. Phys.} B}
\def\PLB{{\em Phys. Lett.}  B}
\def\PRL{\em Phys. Rev. Lett.}
\def\PRD{{\em Phys. Rev.} D}

\def\be{\begin{equation}}
\def\ee{\end{equation}}
\def\bea{\begin{eqnarray}}
\def\eea{\end{eqnarray}}

\begin{document}

\title{
\vspace{-1.5cm}
\begin{flushright}{\normalsize RU-98-39}\\
\end{flushright}
\vspace*{1cm}
CHIRALITY ON THE LATTICE$~{}^*$}

\author{H. NEUBERGER}

\address{Department of Physics and Astronomy, Piscataway,
NJ 08855, USA\\E-mail: neuberg@physics.rutgers.edu}


\twocolumn[\maketitle\abstracts{ 
During the last several years a
 non-perturbative formulation of exact chiral
symmetry on the lattice has been developed. I shall
outline the main ideas of these developments 
and discuss prospects for the future.
The focus will be on the basic concepts enciphered in 
a new jargon consisting of terms like
``infinite number of fermions'', ``domain wall
fermions'', ``the overlap'', ``the Ginsparg-Wilson relation''.
Technical details will be omitted.}]

\section{Introduction}
\footnotetext{${}^*$ Talk delivered at ICHEP'98, UBC, Vancouver, B.C., 
Canada, July 23-29, 1998.}
As far as we know the basic constituents of matter are chiral fermions. 
Their interactions are described by an effective theory, the minimal
standard model. This model is experimentally
established at relatively long scales. 
The scales where field theory will cease being an adequate 
description of Nature are much shorter. Thus, there is a large range
of scales where field theory is reliable. A complete theory would
contain all scales. We are led \cite{hbn_rome} to ask: 
Is it plausible for a complete
theory to contain such a large separation of scales, given that
the long scales are chiral?  If we restrict the question to renormalized, 
perturbative field theory, the answer is positive. But, if we 
now consider the same question in a non-perturbative context, the
answer is less clear. 

To simply ignore the
non-perturbative aspects of the question would be wrong: Non-perturbative
considerations have proven to be relevant before, for example
providing an intrinsic upper bound on the Higgs mass (the triviality
bound) \cite{neub_dallas}. Chiral gauge theories pose difficulties to
non-perturbative analysis. The most famous of these is
encountered in lattice field theory. For a long time it was believed
that chiral gauge theories cannot be defined in any reasonable way
on the lattice \cite{smit_capri}. Developments over the last
six years are poised to falsify this belief.  

The main physics question is then what can be learned from the difficulties
on the lattice and their resolution: Are the problems general
or just a specific lattice quirk? Does the recent resolution point
to a generic structure, relevant also off the lattice, in the real world?
Most experts would agree that the problem is not just a lattice quirk;
nevertheless, most field theorists work in the continuum, and implicitly
assume otherwise. Accepting that the problem is generic 
we would expect a resolution that is not limited to the lattice. 
The recent resolution indeed is not.

In a nutshell, it works by postulating an infinite number of extra
heavy fermions per unit four-volume. All the extra fermions
are as heavy as we wish. Their masses can be kept naturally at
energies many orders of magnitude above the typical energy-momentum
scales of the chiral gauge theory governing the long scales. 
They can be integrated out fully, leaving a consistent low energy
chiral theory. 
The required infinite number of fermions is suggestive of one or more
extra dimensions, beyond the four we know. It is tantalizing that
there exist other speculations about the fundamental laws of Nature
that also use extra dimensions. 

This solution was developed in collaboration with Narayanan
\cite{neub_nar_plb1,neub_nar_npb1,neub_nar_npblong}.
We started to work on the problem in response to 
independently conceived ideas that appeared in two papers. 
The most widely known paper \cite{kaplan} is by Kaplan 
and has roots in earlier
work \cite{callan_harvey} by Callan and Harvey. 
The other paper \cite{frolov_slavnov} was authored by Frolov
and Slavnov. These two papers seemed very different but Narayanan and I
identified a common mechanism. We pursued these insights to 
a concrete realization. Recently it has become clear that some 
germs of the idea can be found
in an earlier paper by Ginsparg and Wilson \cite{ginsp_wils}. 

During the last year many more lattice field theorists have started 
working on implementations of exact chiral symmetry on the lattice. 
Progress has been particularly rapid for vector-like field theories, 
where the exact chiral symmetries are global. This is important
for QCD. Although in QCD the quarks are massive, there are
chiral symmetries that are approximate and their incorporation
in effective descriptions of the low energy properties of QCD
has been extremely successful phenomenologically. Thus, numerical
work on lattice QCD would benefit enormously from a practical
lattice formulation of exact global chiral symmetry. 

\section{Different views of infinite numbers of fermions}

Callan and Harvey were looking for a physical setup to find
connections between anomalies in dimensions that differ by one
or two units. They considered a gauge theory in five dimensions
with a mass term that depended on the fifth dimension going
monotonically from a positive value to a negative one. 
The five dimensional free
Dirac equation has solutions that propagate with the speed of light
along the wall but are exponentially suppressed in directions
perpendicular to the wall. These solutions also turn out to be
unpaired eigenstates of the $\gamma_5$-matrix. They represent chiral
fermions confined to the four dimensional domain wall located at
the point $x_5$ where the mass vanishes. The effective action of the fermions
on the wall is chiral, so can be anomalous.
The associated charge
must leak into the fifth dimension since from the five dimensional
point of view the charge is conserved. 
This charge has to
be allowed to disappear at the $\pm$ infinities in the fifth direction;
one cannot compactify that dimension, as charge would then have
no way of disappearing.
The number of fermions is infinite not
only because the fifth direction is continuous, but, more importantly,
because it is open at the two ends.

Kaplan showed that the Callan-Harvey phenomenon could
be realized on a lattice using a Wilson mass term to create the domain-wall,
the ``defect''. He understood that
the continuity of the fifth direction (in addition to that of the
first four) was not essential. 
But, he abandoned the openness of the ends
of the fifth direction in order to make the whole setup finite. 
This added an anti-domain-wall housing fermions of opposite chirality.
Kaplan speculated that one could devise a gauge action which
could keep the chiral fermions at their respective walls
and the two chiralities would not communicate even 
in the presence of arbitrary gauge fields. So, one still had
a vector-like theory but the two chiral components of the fermions
were ``physically separated'' in the fifth direction and, if
it so happened that there were no anomalies at one of the walls,
one could hope that the walls would dynamically 
decouple. It still could
happen that fermion number would be violated at one of the walls.
Such violations had to be exactly compensated at the other wall - 
the decoupling of the walls was not complete. 
Before Kaplan, Boyanowski et al. \cite{boya} discussed
Callan-Harvey phenomena on the lattice, but their suggestions
were less concrete and attracted no attention. 

Frolov and Slavnov considered the regularization of $SO(10)$
gauge theory with one 16-plet of Weyl fermions.
They introduced an infinite number of heavy fermion
fields of normal and abnormal statistics. 
Their scheme made use of specific $SO(10)$ properties. The choice
of statistics of the fermi fields ensured that the real part of
the effective action induced by the fermions came in with
the required weight, half of that induced by Dirac fermions. 
The ordering
implied by the masses implied that the infinity was one dimensional. 
On the other hand, the gauge fields were just four dimensional.
The entire treatment was in the continuum.   

The papers by Kaplan and Frolov and Slavnov appeared at a time
that the importance of topology induced fermion number
violating processes was recognized. In a bilinear action
different multiplets of fermions do not couple to each other. 
Therefore, one can imagine integrating over the multiplets 
independently. But, in an instanton background, something very strange
must be allowed to happen: a single $\psi$ field may acquire a
nonzero expectation value. This is impossible 
if the kernel of the bilinear form
is a finite square matrix. In the continuum, this is possible because
of a non-zero index. The index
of a linear operator can be understood as a measure of the difference
between the number of rows and the number of columns of a finite
matrix approximating the operator. This difference cannot be frozen:
it must change as the gauge background changes. This can be achieved
only if the kernel is an infinite matrix, in other words, 
the number of fermions is infinite.

Narayanan and I interpreted the papers of Kaplan and Frolov
and Slavnov in a new way. 
Our setup also accommodated the above described instanton
phenomena. We start with \cite{neub_nar_plb1} 
an apparently vectorial gauge theory:
\begin{equation}
{\cal L}_\psi = \bar\psi\gamma_\mu D_\mu \psi +\bar\psi \left (
{{1+\gamma_5}\over 2 }{\cal M} +
{{1-\gamma_5}\over 2 }{\cal M}^\dagger \right )\psi
\end{equation}
The operator ${\cal M}$ acts in a new space and has an analytic
index there (do not confuse this index with the one
associated with instantons): while dim(Ker${\cal M}$)=1,
dim(Ker${\cal M}^\dagger$)=0. This property is radiatively stable,
but can be realized only for dim${\cal M}$=$\infty$. 
The infinite size of ${\cal M}$ and its index make it impossible
to use a standard bi-unitary transformation on the fermi fields
to replace a non-hermitian ${\cal M}$ by a hermitian operator. If in
Kaplan's original approach one made the gauge fields four dimensional
his scheme looked just as Frolov and Slavnov's only the
mass matrices were different. In his case the masses approached
asymptotic values, while Frolov and Slavnov chose them to increase
without bound. But, in both cases one had an index. 

To control
the infinity we interpreted the fifth direction as an 
imaginary time coordinate. So, we had 
to calculate the projectors on the vacua of the many-fermion 
problems associated with asymptotic translations in the fifth direction.
Starting from the $\pm$ infinities, 
two different states propagate inwards, towards the defect. 
The fermion-induced action
of interest is localized at the interface and is contained in
the overlap of the two states - 
hence the term ``overlap'' \cite{neub_nar_npb1}. 

The projectors
are onto rays: thus, while the absolute value of
the overlap is well defined its phase is not. 
This is just right: gauge
invariance is preserved at all steps and because of anomalies 
it has to be impossible to achieve a complete definition 
for arbitrary fermion representations.
We have a $U(1)$ bundle over the space of gauge fields
and the bundle itself can be reduced to the space of gauge orbits.
But, there is no guarantee that reasonable sections of the
bundle can be found: We ended up with a mathematical structure able 
to reproduce the subtle features of chiral
fermions \cite{neub_nar_npblong}. 

Ginsparg and Wilson proposed in 1982 a renormalization
group approach to representing exact global chiral symmetries on the
lattice. The renormalization group is designed to bring out  
quantum versions of scale invariance which are anomalous.
Massless QCD is classically scale invariant and consequently
has exact global chiral symmetry. Therefore, a fixed point 
for massless QCD
should exhibit chiral symmetry and also chiral anomalies. 
For concreteness, Ginsparg and Wilson imagined starting 
the renormalization group iteration from a chirally symmetric
action. The iteration has to break chiral symmetry to make anomalies
possible, but this is the single source of breaking.
The initial symmetry constrains the resulting effective action. 
This action has to satisfy a remnant of chiral symmetry, 
the Ginsparg-Wilson relation. From it, Ginsparg and Wilson derive
the anomalous $U(1)_A$ Ward identity. In footnote 11 they observe that
all non-anomalous Ward identities could be derived in the same manner.
Such identities hold when more than one quark flavor is present ($N_f >1$). 
In footnote 7 they comment that the fixed point action cannot
be bilinear in the fermion fields when $N_f >1$ \cite{eicht_pres}. 

Their paper was forgotten because, in the presence of gauge fields, 
they had no explicit solution to their relation. That solutions should
exist for $N_f=1$ was plausible if one accepted the existence of 
renormalization group transformations with acceptable bilinear fixed points. 
To reach a fixed point an infinite number of iterations are required. 
Each iteration removes a slice of short ranged fermion modes and compensates
by dilating the remainder. 
The chirally symmetric starting point is then separated from the actual 
action by the elimination of an infinite number of fermionic (and 
bosonic) degrees of freedom.

I have mentioned already that the infinite matrix ${\cal M}$
can realize both Kaplan's and Frolov and Slavnov's ideas 
and in its Kaplan-version leads to the overlap. 
In the vector-like context, if one does not insist on explicit
decoupling of left and right Weyl spinors in the action,
the kernel of the bilinear fermion can have an 
unrestricted square shape for any gauge background.
Thus, one can hope for a sequence of kernels
of finite square shape that converge to the infinite kernel
representing the strictly massless case. Such a sequence was 
known since Kaplan: it is nothing but his domain wall
construction, slightly modified \cite{boya,shamir}. Viewed differently, 
it consists of a light Dirac fermion coupled to many heavy 
Dirac fermions \cite{neub_prd} by a mass matrix of seesaw type.
The seesaw suppression produces strictly massless fermions only
after an infinite amplification. But, the mass decreases exponentially
fast.

I have emphasized the infinite number of fermions because it is
a physically appealing picture. Mathematically however, the overlap
construction \cite{neub_nar_npb1} was recast \cite{neub_nar_npblong}
in a way that made no references to anything infinite. As a result,
some technical simplifications were achieved. Moreover, the scheme
is very flexible. This flexibility amounts to freedom to choose
from large classes of lattice $H$-operators. Adapting the form
of $H$ to the specific problem at hand turned out very useful for efficient
numerical simulations in two dimensions and also in analytical work. 
This flexibility will very likely be further exploited in future
applications. 

\section{Finite number of fermions}

In the vector-like case the overlap 
admits an explicit form involving a finite square matrix,
the overlap-Dirac operator \cite{neub_plb_massless}, 
which couples left and right Weyl
components, and is therefore of the type that should obey
the Ginsparg Wilson relation. It should do so because
it was obtained by infinite iteration from an
explicitly chiral starting point. It indeed 
does \cite{neub_prd,neub_plb_gw}, and thus
we finally have an explicit solution to this relation and
the properties established by Ginsparg and Wilson hold.
 
More in line with the original work it has been recently claimed that
a true fixed point to full massless QCD can be replaced
by a classical approximation, called ``a perfect action''
\cite{luscher}, 
and the Ginsparg Wilson relation still holds. The perfect
action is bilinear in the fermions for any number
of flavors, but no explicit expression
is known. An explicit definition is provided only for the
``perfect'' renormalization group transformation and
any fixed point of this map is a ``perfect action''. 
The map contains a minimization step which introduces 
a non-analyticity in the background. Some singularities
are necessary, because of instantons \cite{neub_plb_gw}. 
In the overlap-Dirac operator the singularities are
in direct correspondence with exact zero eigenvalues of a 
finite, local and analytic matrix. 

At present it is unclear whether another class of explicit
solutions to the Ginsparg Wilson relation will be found. 
There are obvious
variations on the overlap itself, but it seems hard to find
something explicit and really new. There also are 
indications \cite{narayanan,chiu_zenkin}
that all acceptable solutions to the Ginsparg Wilson
relation have an overlap ``flavor''. 

The infinite number of fermions imply a dependence only on
the {\sl rays} making up the overlap. On a finite lattice
these rays are points in a $CP({\cal N})$ or
$RP({\cal N})$ space and their Berry phases are the mathematical
vehicle bringing in unavoidable anomalies and the need
for their cancelation \cite{neub_geom}. Anomalies
can show up because the infinite 
number of fermions introduces a lack of determinacy. 
Without it, the regulated chiral determinant would
be a function of gauge fields, rather than a quantity defined up
to phase. The phase freedom is restricted by requiring the
states in the overlap to depend smoothly on the gauge field.
This requirement cannot be made compatible with gauge invariance
if anomalies do not cancel. On the other hand, 
the real part of the induced action
is gauge invariant, no matter what happens with anomalies. 

If one wishes to work within a scheme that makes no reference to
infinite numbers of fermions, the needed rays can be introduced
by hand. However, solutions to the Ginsparg Wilson relation, unlike the
overlap Hamiltonians or true fixed point actions, 
cannot be smoothly dependent on the
gauge fields. Therefore, even if one extracts the relevant subspaces,
it becomes unclear why one should care about Berry phases, given
that the associated operators depend non-analytically on the gauge
fields. To consider Berry phases 
we should allow ourselves to be aware of the smooth 
overlap-Hamiltonians. What distinguishes 
overlap solutions to the Ginsparg Wilson relation is that
these come from analytic Hamiltonians  
and consequently have only singularities of a certain type.
It seems contrived to eliminate these Hamiltonians but keep
the singularity structure. 
In any case, an approach ostensibly based solely 
on the Ginsparg Wilson relation,
but exploiting spectral representations 
of the fermion kernel ends up being equivalent to the overlap construction.

Ginsparg and Wilson knew nothing about the overlap and simply 
assumed the existence of an explicitly chirally
symmetric starting point which had to be left somewhat ill defined. 
It is clear now that one can proceed this way and get a well defined
scheme in the end. But,  
should we really ignore the starting point of the iteration 
and just focus on the relation observed by the fixed point?
I think not: Nature has no reason to first
come up with a Ginsparg Wilson relation and then find a solution -  
this relation is either obeyed or not, but the reason must be elsewhere, 
at a deeper level. 

\section{Main achievements of the overlap}

The overlap has been extensively tested both on and off the lattice.

Explicit computations \cite{neub_nar_npblong,seif} 
in perturbative gauge backgrounds confirmed
that the fermions were indeed 
chiral. In perturbation theory the
generality of the overlap structure was made evident by calculations in 
non-lattice regularization schemes. These continuum
schemes also simplified the needed algebra. 
Perturbation theory was used to compute anomalies, the
vacuum polarization, and to check \cite{yamada} the radiative stability of 
masslessness. It produced both consistent and covariant
anomalies, pinpointed the source for their
difference, and provided insights that could later 
be abstracted outside perturbation theory. Also, even in the vector-like
context \cite{neub_lat98} it was necessary to see how various
``no-go'' theorems were avoided at the level of perturbation theory. 

Numerical work established that instanton effects were correctly 
reproduced \cite{neub_nar_npblong,neub_nar_prl,nar_vra}. 
Rather vexing questions had to be answered.
Was it indeed true that a single fermion could acquire an
expectation value in an instanton background? How could
one have explicit violation of $U(1)_A$ in the vector-like context
without violating any other global axial symmetry?
Eventually it became clear that a fully regulated version was available where
't Hooft's solution to the QCD $U(1)_A$ problem was valid.

For chiral models instanton effects are more dramatic. In two dimensions
it was shown that fermion number violation is reproduced in the
overlap \cite{twod_chiral}. 
The model that was investigated also has composite massless
fermions and provides a simple example where 't Hooft's consistency
conditions are non-trivially respected. Using the overlap, 
this model was simulated numerically, 
and the success of this experiment constitutes the most
subtle test of the overlap to date. The test also shows that 
exact gauge invariance on the lattice is not needed so long
as the model makes sense in the continuum \cite{FNN}. 
A slight breaking of gauge 
invariance amounting to short ranged
correlations between the gauge degrees of freedom $g(x)$ is irrelevant
in the continuum limit where the $g(x)$ 
become independent and decouple from observables.

Non-perturbative anomalies are relatively subtle in continuum physics.
In a non-perturbative setting they should become simpler to understand
than perturbative anomalies. This was shown for the overlap in three
and four dimensions \cite{neub_kik,neub_z2}. 
In both dimensions the mechanism behind 
the anomaly is simple: the non-perturbative anomalies reflect Berry
phase obstructions to choosing global phases of the states making up the 
overlap. These obstructions are directly related to 
overlap-Hamiltonian level crossings over an extended parameter space. 

Three dimensions is a promising area of applications for the overlap
\cite{neub_kik,neub_practical}.
Although there is no chirality in any odd dimensions, there exists an
analogue, and the overlap machinery can be extended to three
dimensions with ease. A particular
set of three dimensional models that were studied by Appelquist
\cite{Appelquist}
and collaborators, admit an overlap formulation.
From it, I derived a three dimensional generalization of the Ginsparg Wilson
relation. Three dimensional models will be instructive. 

\section{Prospects}

The initial apathetic reaction of dominant factions in the lattice community
to the progress on the chiral fermion front has all but disappeared.
The most convincing sign of change are the emerging priority squabbles.

If the overlap is correct and practical difficulties are overcome, 
the way lattice QCD is currently being done will undergo a revolution. 
Currently, a large fraction of the numerical QCD effort, which commands
the bulk of support, manpower and visibility in lattice field theory, is
invested in diminishing and controlling chirality violating effects.
Following initial work on a two dimensional vector-like model \cite{schwinger},
serious studies \cite{vranas_dw} 
of the domain wall truncation of the overlap
have been undertaken 
and consequently the largest computer resources
in the US will be applied 
to the domain-wall-fermion seesaw-approximation 
of the overlap \cite{soni}.

Even more recently, progress has been made also on the direct implementation
of the overlap-Dirac operator on the 
lattice \cite{neub_practical,scri_practical}, relegating all
truncations to their natural place: numerical algorithms. It is too
early to say whether the direct approach or the domain wall one
will ultimately prove more efficient. 
In principle, the direct approach is cleaner.

\section{Summary}

A new and exciting lattice methodology is emerging and, possibly,
we are witnessing a big step forward in numerical QCD. 
At the base of this progress lies a world consisting of an infinite
number of fermions, all but one having very large masses. 
This infinity is fully under control and consequently
can be completely eliminated. 
It is natural to speculate that we have discovered
more than just a trick designed for computers, namely, that we 
have obtained a valuable hint about chirality in Nature.   

\section*{Acknowledgements}

This research was supported in part by the DOE under grant 
\#DE-FG05-96ER40559.

\end{document}